\documentstyle[11pt,paspconf,psfig]{article}

\begin{document}

\title{On the Verge of Coalescence: a Dusty Group of Galaxies}

\author{S. Temporin and R. Weinberger}
\affil{Institut f\"ur Astrophysik, Leopold-Franzens-Universit\"at Innsbruck,
    A-6020 Innsbruck, Austria}

\author{F. Kerber}
\affil{Space Telescope - European Coordinating Facility, ESO, D-85748 Garching, Germany}

\begin{abstract}
We found a new, very compact group of galaxies characterized
by a median projected separation between galaxies of 5.2/h kpc
and a very low velocity dispersion.
All the observed member galaxies show emission line spectra
and a disturbed morphology. 

From these properties it emerges that this compact group is 
in an extremely advanced stage of evolution.

\end{abstract}


\section{Introduction}

Compact groups of galaxies (CGs) are thought to evolve into isolated bright 
ellipticals, as suggested by numerical simulations (e.g. Barnes 1989;
 Athanassoula, Makino \&\ Bosma 1997),
but until now there are very few known examples of elliptical galaxies
whose properties are suggestive of a remnant of 
a merging process of the most luminous members of CGs (Mulchaey 
\&\ Zabludoff 1999; Ponman et al. 1994).
This is why it is very important to have the opportunity to study a CG
in an advanced stage of evolution, that means on the way to coalesce into 
a field elliptical galaxy.

We think to have found such a kind of link between the CG-stage and 
the final product of its evolution: CG1720-678 was identified as an unusual 
object during a survey of the Palomar Observatory Sky Survey plates 
(Weinberger 1995)
and we have now recognized in it several properties indicating an advanced 
stage of evolution.

The properties designating CG1720-678 as a compact group in the classical 
sense and at the same time characterizing it as an extreme object are the 
following:
\begin{enumerate}
\item Extreme density: the median projected separation of the group members
is 5.2 $h^{-1}$ kpc ($h$ = H$_0$/100 km s$^{-1}$ Mpc$^{-1}$). As a 
comparison we remember that the most compact Hickson Compact Group (HCG) has
a median projected separation of 6.8 $h^{-1}$ kpc (Hickson 1993).

\item Very low velocity dispersion: $\sigma_{2D}$ = 67 km s$^{-1}$ and 
$\sigma_{3D}$ = 99 km s$^{-1}$, while the median values for HCGs are:
$\sigma_{2D}$ = 200 km s$^{-1}$ and $\sigma_{3D}$ = 331 km s$^{-1}$
(Hickson et al. 1992).

\item The four member galaxies of the group are within 3 magnitudes from its 
brightest galaxy, with m$_V$ ranging from $\sim$ 17.5 to $\sim$ 20.5, as estimated 
by comparison of our V frame with the images on the SERC J and ESO R film copies.

\item Isolation: as the nearest catalogued galaxy is about 1 degree away and 
an inspection of the POSS plates does not reveal the presence of galaxies of
comparable magnitude within 11 times the diameter of the group ($\sim$ 30$^{\prime\prime}$),
the group fulfills the isolation criteria.

\end{enumerate}

\section{Observations and Data Analysis}

Optical imaging and spectroscopy were used to establish the main characteristics 
of CG1720-67.8: a 900 sec exposure in the V band with a resolution of 0.259
arcsec/pixel and a long-slit spectrum with spatial resolution 0.56 arcsec/pixel,
dispersion 2\AA/pixel over the  range 3770 - 7180 \AA, and spectral resolution
$\sim4$ \AA\ were obtained at the Du Pont 2.5m telescope of Las Campanas Observatory
(February 1998);
another long-slit spectrum with spatial resolution 0.26 arcsec/pixel, dispersion
2 \AA/pixel over the range 3850 - 8000 \AA\ and spectral resolution $\sim12$ \AA\
was obtained at the MPI 2.2m telescope in La Silla (May 1997).
After the usual reduction steps, a laplacian filtering procedure (Richter et al.
1991) was applied to the V-frame in order to suppress the 
noise and enhance faint details. The result is shown in Fig. 1.
The spectra, reduced in a standard way and flux-calibrated by means of 
spectrophotometric standard stars, covered the objects 1 to 6 in Fig. 1.

\begin{figure}
\centerline{\hbox{
\psfig{figure=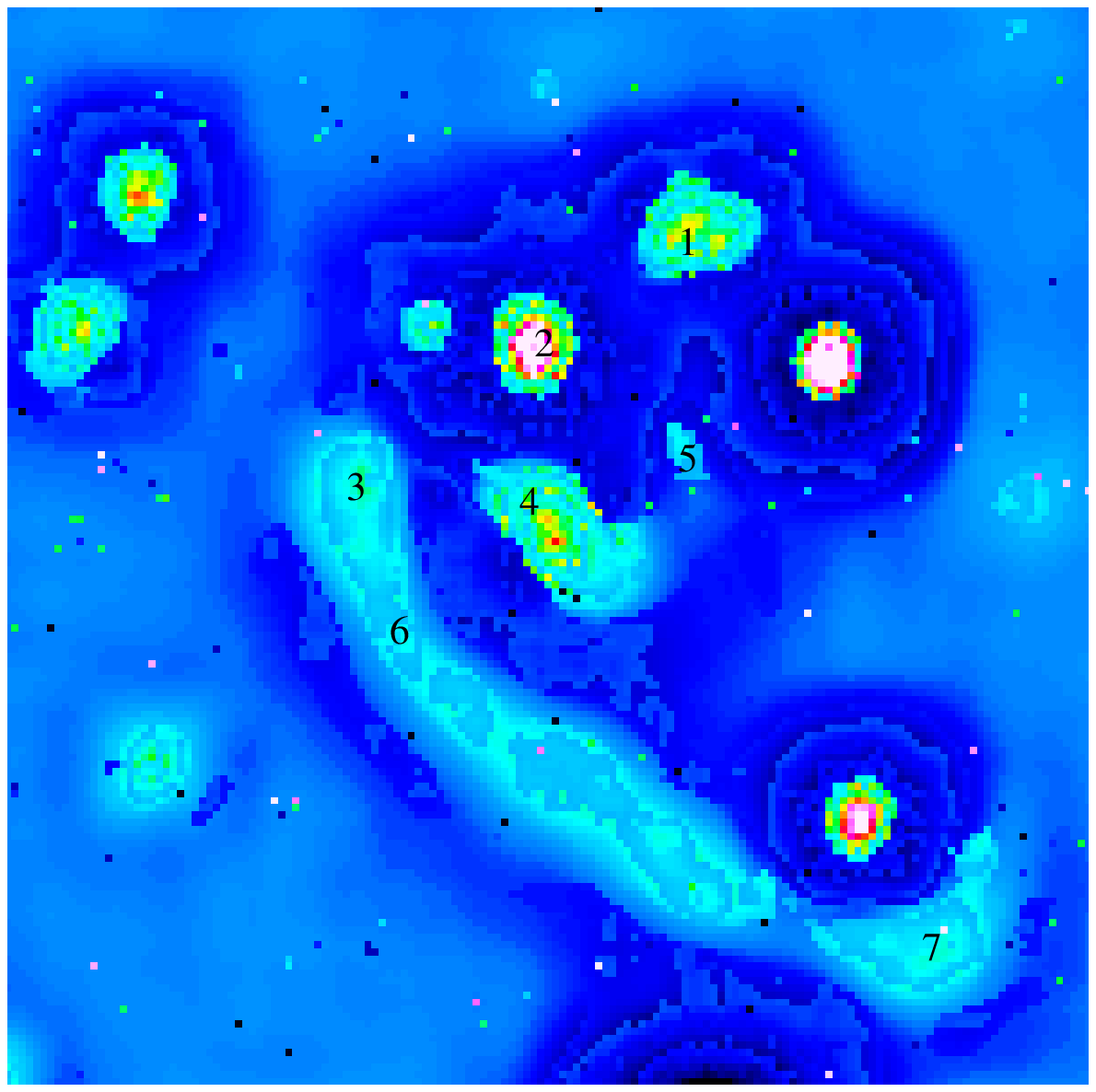,clip=,angle=-90,width=0.4\textwidth}
\psfig{figure=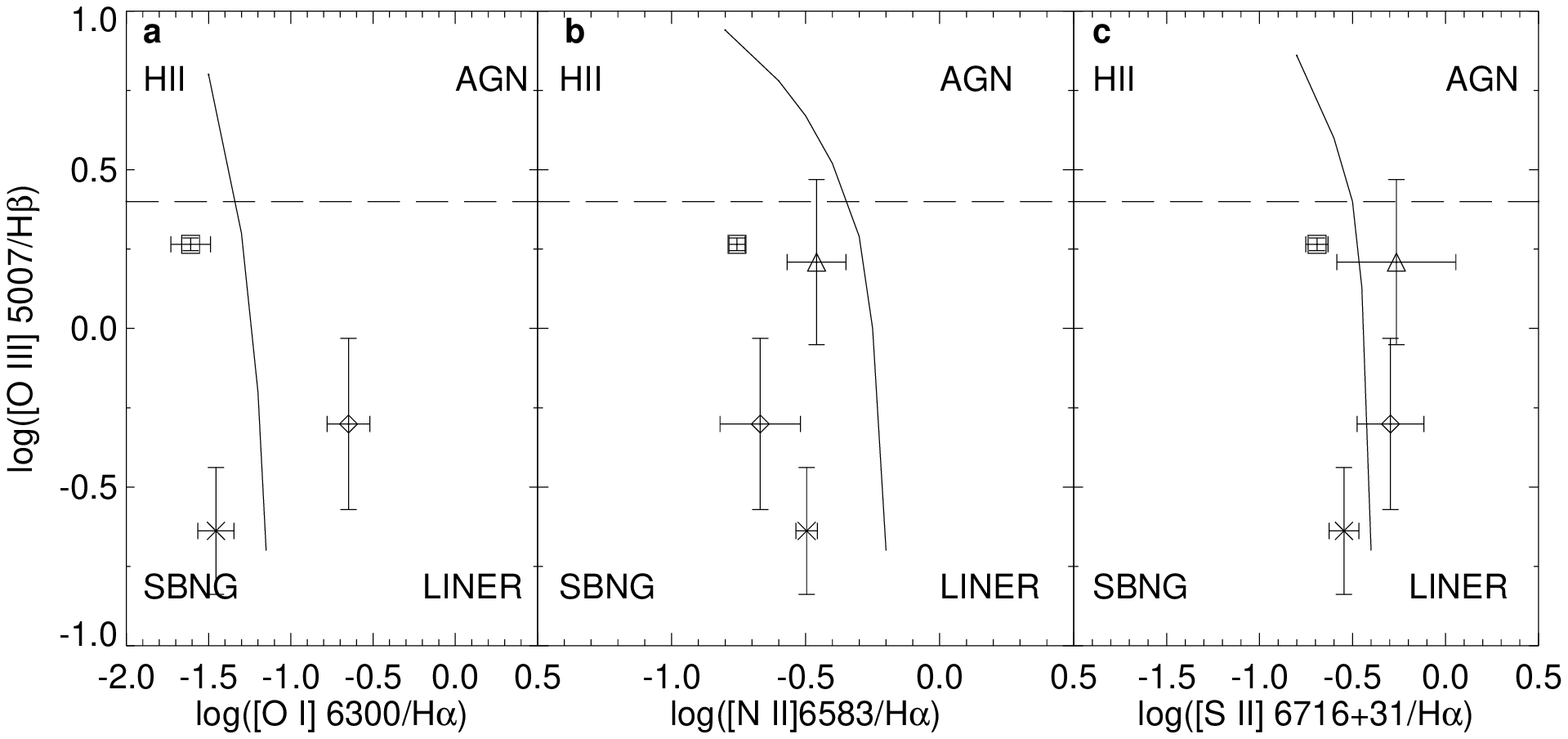,angle=90,width=0.6\textwidth,clip=}}}
\caption{Left: V-band image of CG1720-67.8 after laplacian filtering. Right:
Diagnostic diagrams of the four brightest objects of the group; squares, diamonds,
triangles and crosses represent the emission-line ratios of objects no. 1, 2, 3, and 4 respectively.}
\end{figure}

The three brightest members of the group (no. 2, 4, and 1) may 
apparently be classified as elliptical galaxies, but an analysis of the 
filtered image reveals that two of them have disturbed morphologies: 
no. 4 reveals a possible double nucleus, outer 
structures that could indicate that it is the product of a merging of two disk 
galaxies, and a faint tail ending in an emission knot (no. 5) whose nature is still 
unclear. Object no. 1 shows some internal structure, maybe a double 
nucleus. An outstanding ``arc'' dominates the image of the group and is 
characterized by brightness enhancements at its ends (no. 3 and 7), whose northern one is the 
fourth-brightest member-object. The arc is probably a tidal feature generated 
by interaction and/or merging processes.

Spectra of the objects no. 1 to 5 and of a portion of the arc (no. 6)
were extracted:
all of them show  emission lines and are characterized by blue continua typical
of young stellar populations, except for object no. 2 whose spectrum
suggests a mixed stellar population and could be indicative of a post-starburst 
phase.

Internal extinction estimated from the 
Balmer decrement after having taken into account the effect of stellar
 absorption lines (Table 1), seems to indicate a large amount of obscuration
 (the estimated Galactic foreground extinction is only A$_V$ $\sim$ 0.3).

A classification of the type of activity of the four observed galaxies (Fig. 1, right),
obtained by means of diagnostic diagrams (Veilleux \&\ Osterbrock 1987; 
Coziol et al. 1998), shows that all of these galaxies have a starburst 
nucleus. Star formation rates per unit projected area have been derived from 
H$\alpha$ luminosities (Table 1). The values are comparable or even higher than 
typical SFR of interacting spiral galaxies (Bushouse, 1987), ranging from 
6.11 10$^{-11}$ to 7.6 10$^{-8}$ M$_{\odot}$ yr$^{-1}$ pc$^{-2}$.

\begin{table}
\caption{Estimated non-galactic foreground extinction values (A$_V$), 
and star formation rates (SFR) in units of 10$^{-8}$ M$_{\odot}$ yr$^{-1}$ pc$^{-2}$.}
\begin{center}
\begin{tabular}{ccc}
\tableline
Object & A$_V$ & SFR\\
\tableline 
1 &0.8 $\pm$ 0.1 &15.0\\
2 &1.6 $\pm$ 1.2 &7.3\\
3 &3.2 $\pm$ 1.4 &1.2\\
4 &2.0 $\pm$ 0.4 & 9.9\\
\tableline
\end{tabular}
\end{center}
\end{table}

Puzzling enough, in spite of the high degree of interaction, the enhanced star 
formation activity, and the considerable amount of extinction, there is no
catalogued IRAS source at the position of CG1720-67.8.
After building contour maps of the emission in the four IRAS bands, we found
only a weak emission, at a 4$\sigma$ level on the background, at 60 $\mu$m.
An improved resolution map, obtained after application of the Maximum Entropy 
method, shows that this emission is peaked at the position of the compact group
(Fig. 2).

\begin{figure}
\centerline{\hbox{
\psfig{figure=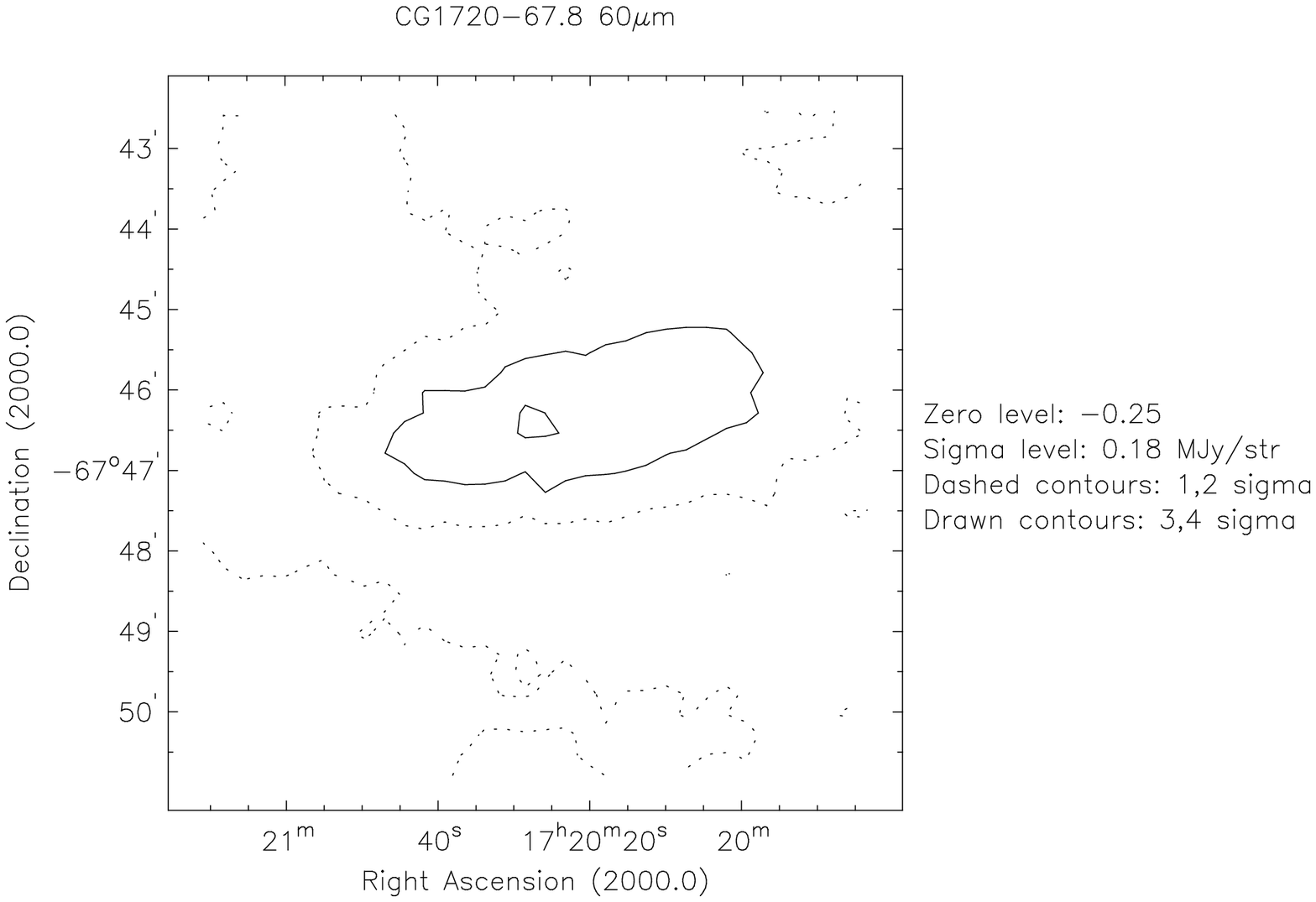,clip=,width=0.4\textwidth}
\psfig{figure=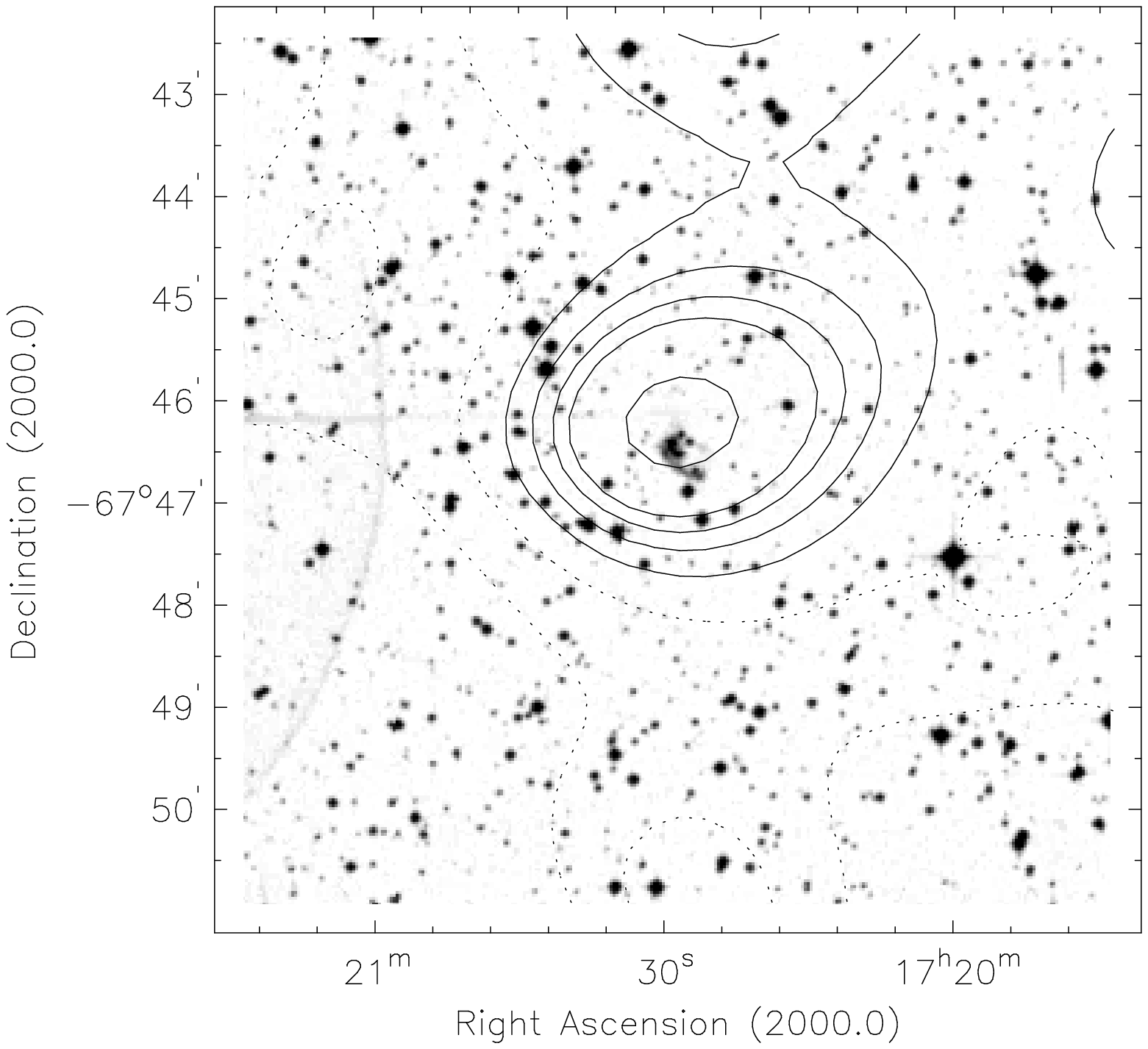,clip=,width=0.27\textwidth}}}
\caption{60 $\mu$m IRAS map (left); Maximum Entropy result overlaid to the 
DSS image of the group (right).}
\end{figure}

\section{Discussion}

CG1720-67.8 is extremely compact, presents a high degree of interaction 
among its members and is dominated by a tidal-like feature. 
Velocity data derived from the spectra of the brightest members reveal a  
very low velocity dispersion. All the observed members have emission-line 
spectra indicating enhanced star formation activity, that we interpret as 
a consequence of the ongoing interactions. Traces of a post-starburst phase 
may be present in the brightest object.
All of these properties lead to suppose that the group is on the verge of 
coalescence.

The most puzzling question is represented by the lack of detected FIR emission 
in spite of the enhanced star formation activity and of the great amount of
(non-Galactic) foreground obscuration. We suggest as a possible explanation 
that the strong extinction is due to cool dust, probably emitting beyond the 
IRAS bands, that could have been blown off far away as a consequence of strong 
galactic winds.

We believe that this group represents an important stage of the CGs' evolution 
and is worthy of further detailed studies.

\acknowledgments
We are grateful to G. Richter for making available to us the Potsdam 
Image Processing System, and to S. Ciroi for providing us with the Maximum Entropy
IRAS map.


\begin{references}
\reference Athanassoula, E., Makino, J., \& Bosma, A. 1997, \mnras, 286, 825
\reference Barnes, J. E. 1989, Nature, 338, 123
\reference Bushouse, H. A. 1987, \apj, 320, 49
\reference Coziol, R., Ribeiro, A. L. B., de Carvalho, R. R., \& Capelato, H. V.
        1998, \apj, 493, 563
\reference Hickson, P. 1993, Astrophys. Lett.\& Comm., 29, 1
\reference Hickson, P., Mendes de Oliveira, C., Huchra, J. P., 
           \& Palumbo, G. C. 1992, \apj, 399, 353
\reference Mulchaey, J. S., \& Zabludoff, A. I. 1999, \apj, 514, 133
\reference Ponman, T. J., Allan, D. J., Jones, L. R., Merrifield, M., McHardy, I. M.,
           Lehto, H. J., \& Luppino, G. A. 1994, Nature, 369, 462
\reference Richter, G. M., Lorenz, H., B{\"o}hm, P., \& Priebe, A. 1991, 
           Astron. Nachr., 312, 345
\reference Veilleux, S., \& Osterbrock, D. E. 1987, \apjsupp, 63, 295
\reference Weinberger R. 1995, \pasp, 107, 58
\end{references}
\end{document}